# Linear time algorithms for Clobber

Vincent D. Blondel, Julien M. Hendrickx and Raphaël M. Jungers

*Department of Mathematical Engineering, Université catholique de Louvain, 4 avenue Georges Lemaitre, B-1348 Louvain-la-Neuve, Belgium*

**Abstract**

We prove that the single-player game clobber is solvable in linear time when played on a line or on a cycle. For this purpose, we show that this game is equivalent to an optimization problem on a set of words defined by seven classes of forbidden patterns. We also prove that, playing on the cycle, it is always possible to remove at least $2n/3$ pawns, and we give a conformation for which it is not possible to do better, answering questions recently asked by Faria et al.

*PACS:*

## 1 Introduction

The combinatorial game Clobber was introduced by Albert, Grossman, Nowakowski and Wolfe in 2002 [5]. It is played with black and white pawns that are placed on a grid, or more generally on the nodes of an undirected graph. White begins and moves a white pawn onto an adjacent black pawn and removes this black pawn from the game. The place formerly occupied by the white pawn is then empty and cannot be occupied anymore. Then Black plays similarly with the black pawns, etc. The player making the last move wins the game. Starting from the following initial conformation (where pawns lie on a line),

●○●○●○●○,

[1] The research reported here was partially supported by the "Communauté francaise de Belgique - Actions de Recherche Concertées", by the EU HYCON Network of Excellence (contract number FP6-IST-511368), and by the Belgian Programme on Interuniversity Attraction Poles initiated by the Belgian Federal Science Policy Office. The scientific responsibility rests with its authors. Julien Hendrickx and Raphaël Jungers are FNRS fellows (Belgian Fund for Scientific Research).

the first play of White could be ●○●○●○○ . Black could follow with ●○●● ○○ , after which White could win by playing ○ ●● ○○ as Black would then be unable to play (It can be proved that with this initial alternate conformation of length 8, the first player can always win). This game has received increasing attention for the last years, and it is for instance possible to play Clobber on the web [1].

We study here solitaire Clobber, which is a single-player version of this game, introduced by Demaine et al. [2].

There are two ways to define this one player version. In one version, the player is forced to play alternatively with white and black pawns as in the 2-player game.

In a second version, the player can move a black or a white pawn at each turn.

In both versions of the game, the goal is to remove as many pawns as possible. The smallest number of remaining pawns that can be obtained from a particular initial conformation is called the *reducibility value* of this conformation. Little is known about these games except that both are difficult. More precisely, in both cases it is NP-complete to determine the reducibility value of a given initial conformation even if the underlying graph is a simple grid [2,3]. We focus on the second version of the solitaire game, where the player can chose at each turn the color he plays. In the sequel we simply refer to the second version as "solitaire Clobber".

In [3] Faria et al. prove that the reducibility value of an initial conformation on a line ($1 \times n$ grid) can be computed in quadratic time and they ask if an algorithm with a smaller complexity exists. From their method, they deduce a cubic time algorithm to compute the reducibility value of conformations on a cycle, which they conjecture to be never larger than $n/4 + O(1)$. We show in this paper that the reducibility value of a conformation on both the line and the cycle can be computed in linear time by an algorithm which also provides a sequence of moves attaining this value. We also exhibit a class of conformations on a cycle whose reducibility value is $n/3+O(1)$, thus disproving the conjecture of Faria et al. We then prove that this is the largest possible reducibility value for cyclic conformations.

The paper is organized as follows: In Section 2 we reformulate Solitaire Clobber on a line as an optimization problem over words. We give then in Section 3 an algorithm that solves this problem in $O(n)$ operations and explicitly provides an optimal strategy for the given conformation. Finally in Section 4 we adapt this algorithm to the cyclic case and prove that the worst reducibility value on a cycle is $n/3 + O(1)$.

## 2  Reformulation as an equivalent optimization problem

In this section we formulate solitaire Clobber on the line as an optimization problem on a finite set of words. For a given conformation, we construct a set of words that represent games on this conformation. The optimal strategy is obtained by choosing the word in this set that minimizes the occurrence of a special letter. We denote by $c$ the binary word representing an initial conformation, with $c_i = \bullet$ or $\circ$, and by $n$ the length $|c|$ of this initial conformation. We also denote by $c_{[i,j]} : i \leq j$ the subword $c_i \ldots c_j$. Before defining our set of words, we show that some moves, although allowed by the rules of Clobber, are never needed in the game.

**Proposition 1** *For each game of Clobber, there is another game in which the same number of pawns are removed and such that no pawn having already moved is taken by a pawn that has not moved yet.*

**PROOF.** Suppose that a pawn $A$ having not moved yet takes a pawn $B$ having already moved. For this to be possible, $A$ and $B$ must have different colors and be adjacent before the move. And since $B$ has already moved, it is adjacent to no other pawn as its previous position must be empty, as for example in $\bullet\circ AB\ \bullet$. So if $A$ takes $B$ it becomes isolated (adjacent to two empty positions) as in $\bullet\circ\ A\ \bullet$, and never interacts with another pawn anymore. Thus, by having $B$ taking $A$ instead of the converse as in $\bullet\circ B\ \ \bullet$, one does not remove any (and often add some) possible further interaction or pawn removal. ∗ ∗ ∗

We therefore assume in the sequel that *a pawn having already moved is never taken by a pawn that has not moved yet.*

By the rules of solitaire Clobber, a pawn can only move from an occupied position to an adjacent occupied position, and an empty position thus never gets re-occupied. Suppose that at some point of the game a pawn moves from $i$ to $i+1$ (resp. from $i+1$ to $i$). The position $i$ (resp. $i+1$) is then empty and no pawn can move from or to it. As a consequence, no pawn can cross the separation between $i$ and $i+1$ anymore. This proves the following Lemma.

**Lemma 2** *A separation between two adjacent positions is crossed by at most one pawn.*

Consider now a game of solitaire Clobber starting on an initial conformation $c$. Based on the lemma above, we associate a word $w$ of length $|w| = n - 1$ to the game by associating a letter to each separation, depending on whether or not it initially separates pawns of different colors, and whether or not it gets crossed by a pawn during the game. A letter $s$ is used for $w_i$ when the two initial pawns have the *same* color: $c_i = c_{i+1}$, and a letter $a$ is used when the colors *alternate*: $c_i \neq c_{i+1}$. We add a right arrow: $\rightarrow$ over the letter if the separation gets crossed by a pawn coming from $i$ to $i+1$, and a left arrow: $\leftarrow$ if it gets crossed by a pawn coming from $i+1$ to $i$. If it does not get crossed, we just add a bar $-$. To each game starting with $n$ pawns corresponds thus a unique word $w \in \Sigma^{n-1}$, with $\Sigma = \{\bar{a}, \bar{s}, \vec{a}, \vec{s}, \overleftarrow{a}, \overleftarrow{s}\}$. For example, the following game

$$\bullet\bullet\circ\circ\bullet \Rightarrow \bullet\bullet\circ \ \circ \Rightarrow \bullet\circ \ \ \circ \Rightarrow \circ \ \ \ \circ \tag{1}$$

is associated to the word $\overleftarrow{s}\overleftarrow{a}\bar{s}\vec{a}$. In the sequel, we use $x$ as a "don't care" letter. The expression $\vec{x}\vec{s}$ denotes for instance both $\vec{s}\vec{s}$ and $\vec{a}\vec{s}$.

Not every word of $\Sigma^*$ corresponds to a valid game of solitaire Clobber. For example, the word $\vec{s}\vec{s}\vec{s}$ would correspond to a game where all pawns have the same color and where nevertheless one would take the three others, which is impossible. We say that a word $w$ is *admissible for an initial conformation $c$* if it can be associated to a valid game on this initial conformation. By extension, we say that a word is *admissible* if it is admissible for at least one initial conformation. Since black and white pawns have symmetric roles and since to a sequence of $s$ and $a$ we can associate two possible initial conformations, it can be seen that an admissible word is always admissible for exactly two opposite initial conformations.

During a game of Clobber, the number of removed pawns is equal to the number of moves, and therefore to the number of separations crossed by pawns. In the corresponding word $w$, the number of removed pawns is thus the number of occurrences of $\vec{x}$ and $\overleftarrow{x}$. So the reducibility value of an initial conformation $c$ is the minimum number of occurrences of $\bar{x}$ in all words admissible for $c$, increased by 1 as those words only have $n-1$ letters. Let us define the seven following classes of forbidden patterns.

(1) $\{\overleftarrow{x}\vec{x}\}$
(2) $\{\vec{s}\overleftarrow{s}\}$
(3) $\{\bar{x}\vec{s}, \overleftarrow{s}\bar{x}\}$
(4) $\{\vec{x}\vec{d}\vec{x}, \vec{x}\vec{d}\bar{x}, \bar{x}\overleftarrow{a}\overleftarrow{x}, \bar{x}\overleftarrow{a}\overleftarrow{x}\}$
(5) $\{\bar{x}\vec{d}\overleftarrow{s}, \vec{s}\overleftarrow{a}\bar{x}\}$
(6) $\{\vec{x}\vec{d}\overleftarrow{a}\overleftarrow{x}\}$
(7) $\{\bar{x}\vec{d}\overleftarrow{a}\bar{x}\}$

In the remaining of this section, we prove the two following theorems.

**Theorem 3** *A word $w$ is admissible if and only if $\bar{x}w\bar{x}$ contains none of the patterns (1)-(7).*

**Theorem 4** *Let $c$ be an initial Clobber conformation. It is possible to reduce*

$c$ to $k$ pawns if and only if there exists a word in $w \in \Sigma^{|c|-1}$ such that

- for each $i$, $w_i \in \{\overrightarrow{s}, \overleftarrow{s}, \overline{s}\}$ if $c_i = c_{i+1}$ and $w_i \in \{\overrightarrow{a}, \overleftarrow{a}, \overline{a}\}$ else.
- $\overline{x}w\overline{x}$ contains none of the forbidden patterns (1)-(7)
- $w$ contains exactly $k-1$ occurrences of $\overline{x}$

Moreover, a sequence of plays reducing $c$ to $k$ pawns can then be deduced in linear time from $w$.

We first prove that the patterns (1)-(7) never appear in a valid word. For this purpose, we need two obvious lemmas.

**Lemma 5** *During a game, no more than two pawns can move onto a certain position $i$. If two pawns move to $i$, one comes from $i-1$ and the other one from $i+1$, and they have different colors. The second to move has the same color as the pawn initially on $i$ and becomes isolated after its move so that it cannot take or be taken by any other pawn.*

**Lemma 6** *In an admissible $w$ where $w_{[j-1,j]} = \overrightarrow{x}\overrightarrow{x}$ (resp. $\overleftarrow{x}\overleftarrow{x}$), the move from position $j$ to $j+1$ represented by $w_j$ takes place after the one from $j-1$ to $j$ represented by $w_{j-1}$. Both are made by the same pawn, whose color is opposite to the color $c_j$.*

**Proposition 7** *An admissible word never contains any of the forbidden patterns (1)-(7).*

**PROOF.** For each class of patterns, we suppose that there is a game of solitaire Clobber for which the associated word $w$ contains the pattern, and show that this leads to a contradiction.

(1) $\overleftarrow{x}\overrightarrow{x}$: This corresponds to a situation where two pawns leave a position, which is impossible as once the position has been left it is empty and never gets re-occupied.

(2) $\overrightarrow{s}\overleftarrow{s}$: Suppose that $w_{[i-1,i]} = \overrightarrow{s}\overleftarrow{s}$, pawns initially on positions $i-1, i, i+1$ have the same color, say white without loss of generality. At some time a pawn $A$ moves from position $i-1$ to $i$ and at some other time an other pawn $B$ moves from position $i+1$ to $i$. It follows from Lemma 5 that $A$ and $B$ are of different colors. Suppose that $A$ is white like the pawn $C$ initially on the position $i$, and that $B$ is black (By symmetry this does not lead to a loss of generality). Then $B$ first takes $C$ and is then taken by $A$. This must be the first move of $A$ for otherwise two pawns (a black one and then $A$) would already have moved to $i-1$ where the initial pawn was white, and it follows from Lemma 5 that $A$ would then be unable to take $B$. The additional rule introduced by Proposition 1 is thus clearly violated as $A$ which has not moved

yet takes $B$ which has already moved.

(3) $\overline{x}\overrightarrow{s}$ (and $\overleftarrow{s}\overline{x}$): Suppose that $w_{[i-1,i]} = \overline{x}\overrightarrow{s}$, and let $A$ and $B$ be the pawns initially located on the positions $i$, and $i+1$. No pawn moves to the position $i$ during the game, therefore the pawn moving from $i$ to $i+1$ is $A$, and it is its first move. Since $A$ and $B$ have the same color, $B$ needs to have first been replaced by a pawn $C$ of another color, which violates the rule introduced in Proposition 1.

(4) $\overrightarrow{x}\overrightarrow{a}\overrightarrow{x}$, $\overrightarrow{x}\overrightarrow{a}\overline{x}$ (and their symmetric versions $\overleftarrow{x}\overleftarrow{a}\overleftarrow{x}$, $\overline{x}\overleftarrow{a}\overleftarrow{x}$): Suppose that $w_{[i-1,i]} = \overrightarrow{x}\overrightarrow{a}$, and without loss of generality that $c_{[i,i+1]} = {\circ}{\bullet}$. It follows from Lemma 6 that the pawn moving from position $i$ to $i+1$ first moved from position $i-1$ to $i$ and is black. Therefore it cannot move to position $i+1$ if this position is not occupied by a white pawn, which initially is not the case. The only way to have position $i+1$ occupied by a white pawn is to have it coming from position $i+2$, which implies that $w_{i+1} = \overleftarrow{x}$, forbidding the patterns $\overrightarrow{x}\overrightarrow{a}\overrightarrow{x}$ and $\overrightarrow{x}\overrightarrow{a}\overline{x}$. A symmetric argument can be applied to forbid the two symmetric versions of these patterns.

(5) $\overline{x}\overrightarrow{a}\overleftarrow{s}$ (and its symmetric version $\overrightarrow{s}\overleftarrow{a}\overline{x}$): Suppose that $w_{[i-1,i+1]} = \overline{x}\overrightarrow{a}\overleftarrow{s}$ and without loss of generality that $c_{[i,i+2]} = {\bullet}{\circ}{\circ}$. During the game, no pawn moves to position $i$, therefore the pawn leaving $i$ for $i+1$ is the black one initially on $i$, which we call $A$. Let $B$ be the pawn moving from $i+2$ to $i+1$, and suppose first that $B$ was initially not on $i+2$. Then by Lemma 5 it must be black to be able to arrive on $i+2$ and move afterwards, and can thus not take nor be taken by $A$, contradicting the fact that $w_{i+1} = \overleftarrow{s}$. So, $B$ is initially on $i+2$, and can only move to $i+1$ once the initial white pawn has been taken by $A$. This is however forbidden by the rule introduced in Proposition 1.

(6) $\overrightarrow{x}\overrightarrow{a}\overleftarrow{a}\overleftarrow{x}$: Consider a word $w$ for which $w_{[i-2,i+1]} = \overrightarrow{x}\overrightarrow{a}\overleftarrow{a}\overleftarrow{x}$. Without loss of generality, we can assume that $c_{[i-1,i+1]} = {\circ}{\bullet}{\circ}$. During the game, two pawns move to position $i$. It follows from Lemma 5 that they have different colors. But applying Lemma 6 to $w_{[i-2,i-1]}$ and to $w_{[i,i+1]}$ shows that they must both be black.

(7) $\overline{x}\overrightarrow{a}\overleftarrow{a}\overline{x}$: Suppose that $w_{[i-2,i+1]} = \overline{x}\overrightarrow{a}\overleftarrow{a}\overline{x}$ and without loss of generality that $c_{[i-1,i+1]} = {\circ}{\bullet}{\circ}$. During the game, no pawn moves to position $i-1$ nor to the position $i+1$. Therefore, the two white pawns initially on these positions both move to position $i$, which by Lemma 5 is impossible. $\qquad *\,*\,*$

In order to prove now that forbidding those patterns is sufficient to characterize the set of admissible words, we need the following lemma.

**Lemma 8** *Two words $w, w'$ are admissible if and only if the word $w\overline{x}w'$ is admissible. As a consequence, $w\overline{x}, \overline{x}w$ and $\overline{x}w\overline{x}$ are admissible if and only if*

*w is admissible.*

**PROOF.** Just notice that $w\overline{x}w'$ represents a solitaire Clobber game which is the juxtaposition of two games represented by $w$ and $w'$. The second part of the result follows from the fact that the empty word is valid and corresponds to an empty game on an initial conformation of one pawn. $***$

**Proposition 9** *Any word $w$ such that $\overline{x}w\overline{x}$ contains none of the forbidden patterns (1)-(7) is admissible.*

**PROOF.**

It follows from a recursive application of Lemma 8 that this proposition just needs to be proved for words containing no occurrence of $\overline{x}$. Consider such a word $w$. Because of the forbidden pattern (1), it must have the form $w = (\overrightarrow{x})^p (\overleftarrow{x})^q$.

We first treat the case where $q = 0, p > 0$, which by symmetry is equivalent to the case where $p = 0, q > 0$. Since pattern (3) $\overline{x}\overrightarrow{s}$ is forbidden in $\overline{x}w\overline{x}$, $w_1 = \overrightarrow{a}$. Then because patterns (4) $\overrightarrow{x}\overrightarrow{a}\overrightarrow{x}$ and $\overrightarrow{x}\overrightarrow{a}\overline{x}$ are forbidden, all other $w_i$ are $\overrightarrow{s}$. Assuming without loss of generality that the first pawn is black, the initial conformation corresponding to $w$ is thus $\bullet\circ^p$, and $w$ can be associated to a game where the first pawn moves $p$ times to the right, taking all the white pawns.

Suppose now that $p, q > 0$. Due to forbidden patterns (3), $w_1 = \overrightarrow{a}$ and $w_{p+q} = \overleftarrow{a}$. This excludes the possibility of having $p = q = 1$ as the pattern (7) $\overline{x}\overrightarrow{a}\overleftarrow{a}\overline{x}$ is forbidden in $\overline{x}w\overline{x}$.

Let us assume that $p, q > 1$. Because the patterns $\overrightarrow{x}\overrightarrow{a}\overrightarrow{x}$ and $\overleftarrow{x}\overleftarrow{a}\overleftarrow{x}$ belong to class (4) and do therefore not appear in $\overline{x}w\overline{x}$, there holds $w_i = \overrightarrow{s}$ for $1 < i < p$ and $w_i = \overleftarrow{s}$ for $p+1 < i < p+q$. There remains to characterize $w_{[p,p+1]}$. Since the patterns (2) $\overrightarrow{s}\overleftarrow{s}$ and (6) $\overrightarrow{x}\overrightarrow{a}\overleftarrow{a}\overleftarrow{x}$ are forbidden, it must be either $\overrightarrow{a}\overleftarrow{s}$ or $\overrightarrow{s}\overleftarrow{a}$. Without loss of generality, we assume that $w_{[p,p+1]} = \overrightarrow{s}\overleftarrow{a}$. Supposing that the first pawn is black, the initial conformation is then $\bullet\circ^p\bullet^{q-1}\circ$, and there is a valid game to which $w$ can be associated. The black pawn initially on the first position begins by taking the $p$ white pawns on its right to obtain a conformation $\bullet^q\circ$. The white pawn on the last position then takes the $q$ black pawns on its left. At the end, $p + q$ pawns have been removed.

There remains to consider the case where $p = 1$ and $q > 1$, and its symmetric version. Since the pattern (5) $\overline{x}\overrightarrow{a}\overleftarrow{s}$ does not appear in $\overline{x}w\overline{x}$, we must have $w_{[1,2]} = \overrightarrow{a}\overleftarrow{a}$. Due again to the forbidden patterns of (4), there holds $w_i = \overleftarrow{s}$ for $2 < i < p+q$, so that $w = \overrightarrow{a}\overleftarrow{a}(\overleftarrow{s})^{q-2}\overleftarrow{a}$. Supposing that the first pawn is black, the initial conformation is $\bullet\circ\bullet^{q-1}\circ$ and a valid game of Clobber can again be associated to $w$, in the same way as when $p,q > 1$.                            ∗ ∗ ∗

Proposition 9 states that a word $w$ such that $\overline{x}w\overline{x}$ contains none of the forbidden patterns is admissible. The juxtaposition of these $\overline{x}$ letters is needed to prevent a pawn from leaving or arriving on the extremal positions from outside the game. Suppose now that a word $w$ is admissible. Then it follows from Lemma 8 that $\overline{x}w\overline{x}$ is also admissible and contains therefore none of the forbidden patterns due to Proposition 7. The absence of forbidden pattern in $\overline{x}w\overline{x}$ is thus a necessary and sufficient condition for admissibility of $w$, which proves Theorem 3. This combined with the fact that the number of remaining pawns at the end of the game depends on the number of occurrences of $\overline{x}$ proves Theorem 4.

## 3  Solving Clobber in linear time

In this section we give a linear time algorithm allowing one to compute the reducibility value of a line solitaire Clobber conformation. A byproduct of this algorithm is an optimal strategy for this particular solitaire Clobber game. We use results of the previous section, so we have to find an admissible word for $c$ that minimizes the number of $\overline{x}$ letters. In order to have a simple algorithm, we first restate Theorem 4 with a simplified alphabet, shorter forbidden patterns and a modified necessary and sufficient condition for admissibility. We define our new alphabet $\Sigma' = \{\overline{x}, \overrightarrow{a}, \overrightarrow{s}, \overleftarrow{a}, \overleftarrow{s}\}$ by merging $\overline{a}$ and $\overline{s}$ into $\overline{x}$. Moreover we replace the forbidden patterns of length four (6) $\overrightarrow{x}\overrightarrow{a}\overleftarrow{a}\overleftarrow{x}$ and (7) $\overline{x}\overrightarrow{a}\overleftarrow{a}\overline{x}$ by shorter ones, namely (6') $\overrightarrow{x}\overrightarrow{a}\overleftarrow{a}$ and (7') $\overrightarrow{a}\overleftarrow{a}\overline{x}$. Note that *their symmetric versions* $\overrightarrow{a}\overleftarrow{a}\overline{x}$ and $\overline{x}\overrightarrow{a}\overleftarrow{a}$ *are still allowed.*

**Theorem 10** *Let $c$ be an initial clobber conformation. It is possible to reduce $c$ to $k$ pawns if and only if there exists a word in $w \in \Sigma'^{(|c|-1)}$ consistent with $c$ containing exactly $k-1$ occurrences of $\overline{x}$ and such that $\overline{x}\overline{x}w\overline{x}\overline{x}$ contains none of the forbidden patterns (1)-(5),(6'),(7'). Moreover, a sequence of moves reducing $c$ to $k$ pawns can then be deduced in linear time from $w$.*

**PROOF.** We begin by proving that patterns (6')-(7') can replace patterns (6)-(7). Forbidding (6') $\overrightarrow{x}\overrightarrow{a}\overleftarrow{a}$ implicitly forbids (6) $\overrightarrow{x}\overrightarrow{a}\overleftarrow{a}\overleftarrow{x}$, so we just need to prove that the former pattern can always be avoided in a valid word without affecting the number of removed pawns. Suppose that there is a valid game for which $w_{[i-1,i+1]} = \overrightarrow{x}\overrightarrow{a}\overleftarrow{a}$ and without loss of generality that $c_{[i,i+2]} = \circ\bullet\circ$. It follows from Lemma 6 that the pawn $A$ making the move from $i$ to $i+1$ is black and therefore that the move takes place after that a white pawn $B$ has come on position $i+1$. Instead of having $A$ taking $B$, one can equivalently have $B$ taking $A$ without violating any rule of Clobber, nor any of the forbidden pattern, and without removing any possible future interaction. So $w_{[i-1,i+1]} = \overrightarrow{x}\,\overrightarrow{a}\,\overleftarrow{a}$ can always be replaced by $\overrightarrow{x}\,\overleftarrow{a}\,\overleftarrow{a}$, and can therefore be forbidden. Observe now that due to this new forbidden pattern the pattern $\overrightarrow{x}\,\overrightarrow{a}\,\overleftarrow{a}\,\overline{x}$ is forbidden, and so are $\overleftarrow{x}\,\overrightarrow{a}\,\overleftarrow{a}\,\overline{x}$ and $\overline{x}\,\overrightarrow{a}\,\overleftarrow{a}\,\overline{x}$ due to the forbidden patterns (1) and (7). As a consequence, the pattern $\overrightarrow{a}\,\overleftarrow{a}\,\overline{x}$ is implicitly forbidden, and we can forbid it explicitly without removing any word corresponding to a valid game. Since $\overrightarrow{a}\,\overleftarrow{a}\,\overline{x}$ is a subpattern of (6), we do not need then to explicitly forbid this latter pattern anymore.

One can see that $\overline{a}$ and $\overline{s}$ never appear separately in the forbidden patterns, but always under the form $\overline{x}$. This is consistent with the intuition, as if the separation between two positions is not crossed by any pawn, the fact that the pawn initially on those positions have same or different colors has no influence on the game. Since $\overline{a}$ and $\overline{s}$ also have exactly the same influence on the number of removed pawns, we can merge them into one letter $\overline{x}$.

Finally, a word $w$ is admissible if and only if $\overline{x}w\overline{x}$ contains none of the forbidden pattern. It follows from Lemma 8 that an equivalent necessary and sufficient condition is the absence of forbidden pattern in $\overline{xx}w\overline{xx}$. ∗ ∗ ∗

Since the forbidden patterns have length three, we would like to represent somehow the subsequences of length three of an admissible word. For this purpose, we use a standard trick in coding theory, which is called higher power coding [4]. The best way to represent this is to construct a directed graph $G(V, E)$ whose vertices are all sequences of *two* letters: $V = \{v_1v_2 : v_1, v_2 \in \Sigma'\}$. Since there are five different letters in our alphabet $\Sigma'$, the graph has 25 vertices. For any word $w$ of length $n$, there is a unique sequence of vertices in the graph: $w_{[1,2]}, w_{[2,3]}, \ldots w_{[n-1,n]}$. Let us consider the complete graph (with self loops allowed) on these 25 vertices. Clearly, one has to forbid edges from any node $v$ to any node $v'$ whenever $v_2 \neq v'_1$. We also remove edges $(v, v')$ such that the word $v_1v_2v'_2$ is forbidden. Any path in the remaining graph which begins and ends at the node $\overline{xx}$ represents a word $\overline{xx}w\overline{xx}$ satisfying the conditions of Theorem 10 and therefore a valid game of solitaire Clobber: the third vertex of the path corresponds to the first two letters of $w$, and each jump from one vertex $v$ to another $v'$ corresponds to the concatenation of the letter $v'_2$. We are now able to state the main theorem of this paper:

**Theorem 11** *The optimal word admissible for an initial conformation $c$ can be found in $\Theta(|c|)$. As a consequence, the reducibility value of $c$ and a sequence of moves attaining this value can be found in linear time.*

**PROOF.** Let $c$ be a given conformation of length $n$. We would like to find a word $w$ admissible for $c$ and minimizing the number of letters $\overline{x}$. By Theorem 10, this is equivalent to finding a path of length $n+1$ in our graph such that

(1) The first and last vertices of the path are $\overline{x}\overline{x}$.
(2) The corresponding word represents a game on the initial conformation $c$. More precisely, the second letter of the $i^{th}$ vertex is of the type $a$ (resp. $s$) if $w_{i-1}$ must be of the type $a$ (resp. $s$), $\overline{x}$ being of both types.
(3) The word minimizes the occurrences of $\overline{x}$; that is, the path minimizes the number of vertices whose second letter is $\overline{x}$.

An efficient way to do that is to weight the edges: an edge has a weight one if it points to a vertex of which the second letter is $\overline{x}$, and a weight zero either. One just needs then to find a path of minimal weight among those satisfying the first two constraints above. This can easily be done in linear time: For $l = 1\ldots n+1$ one just needs to compute and store the paths of minimal weight of length $l$ from the node $\overline{x}\overline{x}$ to each node in the graph. This latter computation can be performed dynamically for growing $l$ with a constant cost at each step. Let indeed $f_l(v)$ be the minimal weight of any path of length $l$ arriving at $v$ and $p_{v',v}$ the weight of the edge connecting $v'$ to $v$. The following recurrence holds
$$f_l(v) = \min_{v'} \left( f_{l-1}(v') + p_{v',v} \right), \qquad (2)$$
where the minimum is taken on all $v'$ such that $(v', v)$ is allowed as $l^{th}$ edge. As the number of edges and of vertices is bounded, each iteration can be done in $O(1)$. $\quad***$

At each position of the path in the graph above, only at most 9 vertices among the 25 of the graph are allowed. Indeed for each $l$ at most 9 vertices satisfy condition (2), i.e. are consistent with the initial conformation. Moreover, the in and out-degrees of a vertex are at most 3 if only edges connecting vertices consistent with the initial conformation are considered. So, equation (2) has to be computed for at most 9 vertices, and each time at most three vertices $v'$ have to be considered.

## 4 Solitaire Clobber on a cycle

In this section we analyze the cyclic conformation for solitaire Clobber, and answer questions recently raised in [3]. We briefly show how the results of the previous sections generalize to the cycle, and we give then a linear time algorithm for computing the reducibility value of a cyclic solitaire Clobber game. We then prove that the maximal reducibility value on a cycle is $n/3$.

**Theorem 12** *The reducibility value of a solitaire Clobber game on a cycle is computable in linear time.*

**PROOF.** Let us consider a conformation for which there are black and white pawns. Without loss of generality, let us suppose that $c_{[1,2]} = $ ●○. Observe that if the black pawn initially on $c_1$ moves to the right, the problem becomes equivalent to a solitaire Clobber on the line with conformation $c$. Also, if this pawn moves to the left, the problem becomes equivalent to a solitaire Clobber on the line with conformation $c_{[2,n]}c_1$. The same analysis can be done with the white pawn initially on $c_2$. Now if neither of these pawns moves, at least one of them has to be removed from the game at the end, for otherwise the number of removed pawn could be increased by having one of them taking the other one. This is only possible if the first black pawn at the right of $c_2$ moves backward to $c_2$, or conversely if the first white pawn at the left of $c_1$ moves to $c_1$. In both cases, the game then becomes equivalent to a solitaire Clobber on the line. So the solitaire Clobber on the cycle is equivalent to one of the six situations presented above. One just has then to compute the reducibility value of these six games on the line, and the smallest one is the reducibility value of the cycle. ∗ ∗ ∗

We now analyze the worst possible reducibility value on the cycle. It is proved in [2] that the reducibility value of the so called checkerboard conformation $(●○)^{(n/2)}$ on the line is $n/4 + O(1)$. One can check that this remains true when this conformation lies on a cycle. In [3], Faria et al. ask whether this is the worst (non-trivial) reducibility value that can be obtained on a cycle. We provide a negative answer to this question by showing an initial conformation on the cycle for which the reducibility value is $n/3$, and prove then that no higher value can be obtained. The conformation is simply the juxtaposition of the following pattern: ●○○ .

**Theorem 13** *The cyclic conformation ●○○●○○ ... ●○○ has a reducibility value of $n/3$, where $n$ is the number of pawns in the initial conformation.*

**PROOF.** As done in Section 2 for the line, we can associate to each valid game of solitaire Clobber on the cycle a (cyclic) word of $\Sigma^n$. For exactly the same reasons as in Proposition 7, none of the patterns (1)-(7) appears in such a cyclic word if it can be associated to a valid game.

A word representing a valid game of Clobber solitaire on the cycle for this initial conformation must be of the form $aasaas\ldots aas$. We split this word into "cells" $aas$, and prove that on average there is at least one letter $\bar{x}$ in each cell. More precisely, we show that between two cells containing no letter $\bar{x}$, there must be a cell with two such letters. This implies that the reducibility value is at least $n/3$ because there are at most $2n/3$ arrows, and thus at most $2n/3$ pawns removed. Suppose that a cell has no $\overline{x}$ letter, by applying the rules of Proposition 7, the cell must be $\overrightarrow{a}\overleftarrow{a}\overleftarrow{s}$ or $\overrightarrow{a}\overrightarrow{a}\overleftarrow{s}$ and has thus a $\overleftarrow{s}$ as last letter. This latter fact, combined with the forbidden pattern (3) of Proposition 7, implies that the next cell must be $\overleftarrow{a}\,\overline{as}$ or $\overleftarrow{a}\,\overline{a}\overleftarrow{s}$. So, either this cell has two $\overline{x}$ letters, or it has only one $\overline{x}$ letter, but then it also has $\overleftarrow{s}$ as last letter. This implies that between two cells containing no $\overline{x}$, there must be at least one cell containing two letters $\overline{x}$, and therefore that the reducibility value is at least $n/3$. Observing that moving each black pawn twice to the right hand side effectively removes $2n/3$ pawns is then sufficient to achieve the proof.∗∗∗

We end this section by showing that $n/3 + O(1)$ is the worst possible reducibility value on a cycle with $n \geq 3$, except of course in the trivial case of a monochromatic cycle. Our proof is constructive as it provides a simple strategy allowing the removal of at least $2n/3 - O(1)$ pawns, for any solitaire Clobber game on a cycle. The strategy is summarized as follows: cut the cycle in such a way to obtain a line clobber game ending with ○●, and remove iteratively a maximum of pawns at the beginning of the line, letting at each step one pawn isolated.

We claim that it is always possible to remove at least two pawns at every step, except perhaps for the last one. Indeed if the line begins with ○○, take the next black pawn (which must exist since the line ends with ○●) and remove all the beginning white pawns (there are at least two of them). The situation is identical if the line begins with ●●. If the line begins with ○●, and if these are not the last pawns, write the beginning of the line as ○●…●○…○●. It is possible to remove all these pawns but the first one, by first moving the last (black) one to the left until it sticks to the black series, and then removing this black series with the first pawn. Still, more than two pawns are removed. If they are the two last pawns, one of them can be removed. Finally, if the line begins by ●○, either the same argument can be applied as for ○●, or only two black pawns remain in the line, which is of the form ●○…○●. In the latter case, which can happen only once during the game, all pawns can be removed but two. So for each pawn that is not removed, there are always at least two pawns that are removed, except at the end of the game where one or two pawns can be left while at least one is removed. The number of remaining pawns is thus at most $n/3 + O(1)$. This together with Theorem 13 shows that the worst possible non-trivial reducibility value on the cycle is $n/3 + O(1)$.

## 5   Conclusions

In this paper, we have shown that playing solitaire Clobber on a line or a circle can be viewed as an optimization problem on a set of words with forbidden patterns. This equivalence has allowed us to design a linear-time algorithm computing the reducibility value of an initial conformation and providing an optimal strategy, for both the cycle and the line. This linear complexity is clearly optimal as reading the initial conformation already requires $n$ operations. Using this forbidden pattern approach and a simple alternative algorithm, we were also able to easily prove that the maximal possible reducibility value on a cycle is $n/3 + O(1)$.

It remains open to determine to which extend our approach can be applied to other topologies. On a general graph, one can build a "word" corresponding to a game of solitaire Clobber by assigning a symbol to every edge, in the same way as we do for each separation between positions. This is however most probably ineffective in the general case, as computing the reducibility value is known to be NP-complete for general graphs [3].